\documentclass[useAMS,usenatbib]{mn2e} \input{epsf}

\newif\ifAMStwofonts 
 
\usepackage{longtable}
\usepackage{lscape}

\def\lesssim{\mathrel{\hbox{\rlap{\hbox{\lower4pt\hbox{$\sim$}}}\hbox{$<$}}}}
\def\gtrsim{\mathrel{\hbox{\rlap{\hbox{\lower4pt\hbox{$\sim$}}}\hbox{$>$}}}}
\def\msun{${\rm M}_{\odot}$~}
\def\msol{${\rm M}_{\odot}$}

\def\l_lsun{$\log{L/\rm L_{\odot}}$~}
\def\masa_msun{$M/ \rm M_{\odot}$~}

\def\m_mstar{$M/M_{*}$~}

\def\aap{A\&A}

\def\apj{ApJ}
\def\apjl{ApJ}
\def\apjs{ApJS}
\def\mnras{MNRAS}

\def\araa{ARA\&A}
\def\physrep{Phys. Rep.}

\def\nat{Nature}

\title[Evolution of low - mass CBSs]{Evolution of low mass close binary 
systems with a neutron star: its dependence with the initial neutron star mass}

\author[M.~A.   De   Vito \& O.~G. Benvenuto] {
M.~A.  De Vito$^{1,2}$\thanks{Member of  the Carrera del Investigador
Cient\'{\i}fico, Consejo Nacional de Investigaciones Cient\'\i ficas y 
T\'ecnicas (CONICET). Email: adevito@fcaglp.unlp.edu.ar}, 
O.~G. Benvenuto$^{1,2}$\thanks{Member of  the Carrera del Investigador
Cient\'{\i}fico, Comisi\'on de  Investigaciones Cient\'{\i}ficas de la
Provincia    de    Buenos    Aires    (CIC). Email:
obenvenuto@fcaglp.unlp.edu.ar}
\\
$^{1}$ Facultad de Ciencias Astron\'omicas y Geof\'{\i}sicas, Universidad
Nacional de La Plata (UNLP), \\ Paseo del Bosque S/N, B1900FWA, La
Plata, Argentina\\
$^{2}$ Instituto de Astrof\'{\i}sica de La Plata, IALP,
CCT-CONICET-UNLP, Argentina }

\begin{document}

\date{September 2}

\pagerange{\pageref{firstpage}--\pageref{lastpage}} \pubyear{2008}

\maketitle \label{firstpage}

\begin{abstract}

We  construct  a set  of  binary  evolutionary  sequences for  systems
composed by  a normal, solar  composition, donor star together  with a
neutron star. We consider a variety of masses for each star as well as
for the initial orbital period corresponding to systems that evolve to
ultra~-~compact    or   millisecond   pulsar~-~helium    white   dwarf
pairs. Specifically,  we select  a set of  donor star masses  of 0.50,
0.65,  0.80, 1.00,  1.25,  1.50,  1.75, 2.00,  2.25,  2.50, 3.00,  and
3.50~\msol, whereas for the accreting neutron star we consider initial
masses values  of 0.8,  1.0, 1.2, and  1.4~\msol. Because  the minimum
mass for a proto~-~neutron  star is approximately 0.9~\msol, the value
of  0.8~\msun  was selected  in  order to  cover  the  whole range  of
possible initial  neutron star masses. The  considered initial orbital
period interval ranges from 0.5 to 12 days.

It is found  that the evolution of systems,  with fixed initial values
for the orbital period and the  mass of the normal donor star, heavily
depends upon the mass of the  neutron star. In some cases, varying the
initial   value  of  the   neutron  star   mass,  we   obtain  evolved
configurations  ranging  from   ultra~-~compact  to  widely  separated
objects.

We also  analyse the dependence of  the final orbital  period with the
mass of the white dwarf.  In agreement with previous expectations, our
calculations  show that  the final  orbital period~-~white  dwarf mass
relation  is  fairly insensitive  to  the  initial  neutron star  mass
value. A new  period~-~mass relation based on our  own calculations is
proposed,  which is  in  good agreement  with period~-~mass  relations
available in the literature.

As consequence of considering a  set of values for the initial neutron
star  mass, these  models  allow finding  different plausible  initial
configurations  (donor  and neutron  star  masses  and orbital  period
interval) for some of the best  observed binary systems of the kind we
are interested in here. We  apply our calculations to analyse the case
of PSR~J0437-4715, showing that there is more than one possible set of
initial parameters (masses, period  and the fraction $\beta$ of matter
accreted by the neutron star) for this particular system.

\end{abstract}


\section{Introduction} \label{sec:intro}

During past  years binary  radio pulsars have  been detected  more and
more  often.   Now  we are  aware  of  the  existence of  141  pulsars
belonging to binary systems (ATNF Pulsar Catalogue: www.atnf.csiro.au;
Manchester et al.   2005). For these objects, both  estimations of the
median  mass of  the  companion (assuming  an  orbital inclination  of
$60^o$) and orbital period of  the binary system are available.  If we
restrict ourselves to low mass  companions ($M < 0.35$~\msol), we find
about  100 objects;  approximately half  of them  located  in globular
clusters. Among this  group of binary systems, we  are interested on a
subgroup composed  by a neutron star  (NS) and a low  mass white dwarf
(WD). Presumably, these  objects have a helium rich  interior and will
be referred to as HeWDs.

\begin{centering}
\begin{table*}
\caption{\label{table:shapiro-delay-data}  The  close  binary  systems
composed by  a millisecond pulsar and a  low mass WD for  which it has
been  possible to  detect the  Shapiro  delay effect  and measure  the
masses of both  components.  All these systems belong  to the Galactic
plane population.  From left to right,  the Table presents  the name of
the pulsar,  its spin  period, the WD  and pulsar masses,  the orbital
period and the relevant reference.}
\begin{tabular}{cccccc}
\hline
\hline
Name & $P_p$ & $M_{WD}$ & $M_{NS}$ & $P$ & Reference \\
     & [ms]  & [\msol]  & [\msol]  & [d] &           \\
\hline PSR~J0437-4715 & 5.757  & $0.236 \pm 0.017$         & $1.58 \pm 0.18$        & 5.741  & van Straten et al. (2001) \\
\hline PSR~J1713+0747 & 4.57   & $0.28 \pm 0.03$           & $1.3 \pm 0.2$          & 67.825 & Splaver et al. (2005) \\
\hline PSR~B1855+09   & 5.362  & $0.258^{+0.028}_{-0.016}$ & $1.50^{+0.26}_{-0.14}$ & 12.327 & Kaspi et al. (1994) \\
\hline PSR~J1909-3744 & 2.947  & $0.2038 \pm 0.0022$       & $1.438 \pm 0.024$      & 1.533  & Jacoby et al. (2005) \\
\hline \hline
\end{tabular} \end{table*} \end{centering}

Remarkably,  for some of  the above~-~referred  binary systems  it has
been possible  to make reliable  determinations of the masses  of both
components.   This   has  been   possible  taking  advantage   of  the
relativistic effect know as Shapiro delay (see Taylor \& Weisberg 1989
and  references therein).  In  Table~\ref{table:shapiro-delay-data} we
list the main parameters of these binary systems.  Apart from the data
included there, Nice et al.  (2008) have reported further observations
of  the binary  system  containing PSR~J0751+1807.  They improved  the
values of the pulsar mass, finding it to be $1.26 \pm 0.14$~\msun ($68
\%$ confidence)  or $1.26 \pm 0.28$~\msun ($95  \%$ confidence). These
values are  much lower than their  previous claim (Nice  et al. 2005),
specially in connection  with the inferred mass of the  NS.  We do not
included  them  in  Table~\ref{table:shapiro-delay-data}  because  the
value of the WD mass is not yet available.

The formation  mechanism of such  close binary systems (CBSs)  is well
established:  a low mass,  normal star  undergoes Roche  lobe overflow
(RLOF) and  transfers mass  to a NS  companion.  After a  long, stable
mass transfer  episode the  donor (normal) star  has lost most  of its
mass.   In the  non~-~conservative case,  only  part of  this mass  is
accreted by the NS which is  spinned up, allowing it to be detected as
a  millisecond pulsar  (MSP)  while its  companion  (initially a  main
sequence star)  cools down becoming  a WD (see, e.g.,  Bhattacharya \&
van den Heuvel 1991).

Let  us  make a  brief  discussion  of  the binary  evolution  results
available in the  literature related to the objects  we are interested
in. Sarna,  Antipova \& Muslimov (1998) investigated  the evolution of
CBSs to account for the  binary system containing the MSP PSR~J1012+53
and its  low mass companion.   For the initial NS  mass ($(M_{NS})_i$)
they assumed  the ``canonical'' value  of 1.4~\msol.  Ergma,  Sarna \&
Antipova  (1998) made evolutionary  calculations of  low mass  CBSs in
conservative  and  non~-~conservative  cases  considering  donor  star
masses in the range  $1.0$~\msol~$\leq M \leq 1.5$~\msol.  Again, they
set  $(M_{NS})_i=1.4$~\msol.   Tauris   \&  Savonije  (1999)  computed
non~-~conservative evolution  of CBSs with  low mass (1.0~-~2.0~\msol)
donor  stars and  a $(M_{NS})_i=$1.3~\msun  accreting NS.  The initial
orbital periods range was between 2 and 800~d. Besides, they revisited
the orbital  period~-~WD mass  relation ($P~-~M_{WD}$) in  wide binary
WD~-~radio pulsar  systems.  Podsiadlowski, Rappaport  \& Pfahl (2002)
performed a systematic study of  the evolution of low and intermediate
mass   binary   systems.    In   their   calculations   they   assumed
$(M_{NS})_i=1.4$~\msun for  the NS, which  accretes (at most)  half of
the transferred  matter, while donor stars had  initial masses between
0.6 and 7~\msol. The initial orbital periods covered the interval from
approximately 4~hr to 100~d.  Ergma,  \&  Sarna (2003)  constructed
binary  evolution  sequences  to   account  for  the  observed  binary
parameters    for     PSR~J1740-5340.     Again,    they    considered
$(M_{NS})_i=1.4$~\msol.   Nelson   \&  Rappaport  (2003)  investigated
possible  scenarios for  accretion~-~powered  MSPs in  ultra~-~compact
binaries.  They   calculated  a  large  set   of  evolutionary  tracks
corresponding  to  different  donor  masses and  degrees  of  chemical
evolution at  the onset of mass  transfer. The range  of initial donor
masses was between 1.0  and 2.5~\msun and $(M_{NS})_i=1.4$~\msol. They
assumed a fully non~-~conservative mass transfer case. Benvenuto \& De
Vito (2005) computed the evolution  of a set of binary systems leading
to the formation of HeWDs~-~MSP or ultra~-~compact systems considering
diffusion.  They also analysed  possible progenitors  for some  of the
best observed  systems containing a MSP  together with a  low mass WD.
They  set  $(M_{NS})_i=1.4$~\msun  and  $\beta=0.5$  ($\beta$  is  the
fraction  of  transferred matter  accreted  by  the  NS), although  in
fitting the masses  and orbital period of these  systems, they allowed
for lower  values of  $\beta$. Benvenuto, Rohrmann  \& De  Vito (2006)
found a possible original configuration that accounts for the observed
parameters  of PSR~J1713+0747 binary  system. They  computed a  set of
binary evolution  calculations in order to  simultaneously account for
the  masses  of both  stars  and  the  orbital period,  again  setting
$(M_{NS})_i=1.4$~\msol.

In spite  of the  fact that  in most of  theoretical studies  aimed to
explore the evolution  of low mass WD~-~NS binary  systems the initial
mass of the NS has  been set to $(M_{NS})_i= 1.4$~\msol, observational
evidence  presented in  Table~\ref{table:shapiro-delay-data} indicates
that $(M_{NS})_i$ may indeed be lower.   At present we do not know the
value of  the fraction $\beta$.   The only physical limitation  is the
Eddington critical accretion  rate $\dot M_{NS} \leq \dot  M_{Edd} = 2
\times 10^{-8}$~\msol/yr  (where $\dot{M}_{NS}$ is  the accretion rate
of   the   NS).   Usually    $\beta$   is   considered   as   a   free
parameter.  Certainly,  we may  account  for  NS  masses greater  that
1.4~\msun by  setting an initial canonical value  for $(M_{NS})_i$ and
adjusting $\beta$.  However, this is {\it not} possible if observed NS
masses are lower than 1.4~\msun (e.g., the case of PSR~J1713+0747, see
Table~\ref{table:shapiro-delay-data}). This fact induced us to perform
a systematic  exploration of the  evolution of these CBSs  varying the
initial donor (normal) and  accretor (neutron) stars masses (and also,
the initial orbital period). This is  one of the main purposes of this
paper. 

In our models we consider masses for the donor stars in the range from
0.50   to  3.50~\msol,   and   accreting  NSs   with  initial   masses
($(M_{NS})_i$)  of  0.8,  1.0,   1.2,  and  1.4~\msol.  The  range  of
$(M_{NS})_i$   we   propose    for   our   calculations   needs   some
justification. It is  well known that most of  the accurately measured
NS  masses are  near 1.4~\msol.  Also, it  is well  known  (see, e.g.,
Lattimer \& Prakash 2004 for  a recent tabulation; Lattimer \& Prakash
2007) that  the masses  of some  NSs are  well below  that  value.  In
particular,  these are  the cases  of the  NSs in  the  X-Ray binaries
SMC~X-1,  Cen~X-3 and  4U1538-52 that,  following Lattimer  \& Prakash
(2004), have masses of $1.17^{+0.16}_{-0.16}$, $1.09^{+0.20}_{-0.36}$,
and $0.96^{+0.19}_{-0.16}$~\msun respectively.  More recently, van der
Meer et  al.  (2007) have  presented more accurate  determinations for
the masses  of NSs in binary  systems. Specifically, for  the cases of
SMC~X-1    and     Cen~X-3,    the    authors     find    values    of
$1.06^{+0.11}_{-0.10}$~\msun      and     $1.34^{+0.16}_{-0.14}$~\msun
respectively. Notice that  in the case of SMC~X-1,  the NS is somewhat
less massive, but for Cen~X-3, the NS is notably more massive that the
previous determination.

NSs have  both minimum  and maximum mass  limits. The maximum  mass is
unknown,  but lies in  the range  of $1.44$  to $3.$~\msol.  The upper
bound  follows from  causality  arguments (Rhoades  \& Ruffini  1974),
imposing that the speed of sound in dense matter must be less than the
speed  of  light,  whereas the  lower  bound  is  set by  the  largest
accurately  measured pulsar  mass, $1.4408  \pm 0.0003$~\msol,  in the
binary pulsar PSR~1913+16 (Weisberg \& Taylor 2003).

Regarding to the  minimum NS mass value $M_{min}$,  it is important to
remark  that it  is sensitive  to the  equation of  state (EOS)  of NS
matter  at  sub-nuclear densities.  Haensel  et  al. (2002)  calculate
$M_{min}$ for  cold NSs using  two different EOSs.  For non~-~rotating
configurations  they  find  $M_{min}=0.094$~\msun  for  the  SLy  EOS
(Chabanat  et al.  1998)  and $M_{min}=0.088$~\msun  for the  FPS EOS
(Lorenz  et al.  1993). However,  we are  interested in  rotating NSs,
i.e., the  accreting companion  of a donor  star in CBSs.   Haensel et
al.  (2002) performed  accurate calculations  of stationary,  cold NSs
configurations,  rotating uniformly  at $\nu=100$~Hz  and $\nu=641$~Hz
(which  corresponds  to  the  shortest observed  pulsar  period).  The
authors  find  for   SLy  EOS  that  minimum  mass   at  $\nu=641$  is
$0.61$~\msun  and  for  FSP  EOS,  at  the  same  rotation  frequency,
$0.54$~\msol. For  the case  of $\nu=100$~Hz and  SLy EOS  the minimum
mass finding is  of $0.13$~\msol, $\approx 40\%$ larger  than that for
static NSs.

If we consider  newly born proto~NSs, both thermal  (after core bounce
the  proto~-~NS   has  a   temperature  $T  \approx   10^{10}$~K)  and
neutrino~-~trapping  effects  are  large,  and are  found  to  largely
increase  the  $M_{min}$  value  to  $0.9 -  1.1$~\msun  (Goussard  et
al. 1998; Strobel et al.  1999). Thus, if NSs formation corresponds to
a gravitational collapse  event we should expect the  existence of NSs
with   masses    above   the   $M_{min}$    value   corresponding   to
proto~NSs.  $0.9$~\msol. Observational data  supports this  lower mass
limit.

There is  a large  gap between  the values of  $M_{min}$ for  cold and
proto~-~NS  as estimated  from the  different models  presented above.
Still, a  NS may reach mass  values smaller than $0.9  - 1.1$~\msun by
mass loss after becoming a  cold NS\footnote{Notice that the NS spends
only  several seconds  in releasing  most  of its  lepton and  thermal
content to  become a cold NS.}.  This possibility has  been studied by
Blinnikov  et  al. (1984);  Colpi  et  al.  (1991); and  Sumiyoshi  et
al. (1998). However analysing such possibility and its consequences is
beyond the scope of this paper

In  view  of   the  above  discussion,  the  minimum   NS  mass  value
($0.8$~\msol)  considered  in   our  calculations  may  seem  somewhat
low. However,  in any case, in performing  our theoretical experiment,
we  select the  minimum  value  of the  accreting  NS of  $0.8$~\msol,
somewhat less  massive that the minimum presented  by the observations
and for the theoretical calculations  of proto~-~NS, simply to be sure
we are exploring the whole meaningful NS mass interval.

Is well  know that there exist  a somewhat tight  relation between the
mass  and the radius  of the  cores of  low~-~mass giants  (see, e.g.,
Joss, Rappaport \&  Lewis 1987). Then, a $P~-~M_{WD}$  relation can be
derived. This  will be  valid if  the star belongs  to a  close binary
system \textit{and} undergoes RLOF as a giant.  In the calculations to
be  presented below,  some donor  stars  undergo RLOF  as red  giants;
however,  other  experience  RLOF   when  they  are  still  much  more
compact. Thus, we explore the $P~-~M_{WD}$ relation and test the claim
(Rappaport et al.  1995) that it is nearly independent of $(M_{NS})_i$
quantitatively and  in more  general conditions than  those previously
considered.

The   reminder   of  this   paper   is   organized   as  follows:   In
Section~\ref{sec:numerical} we briefly  describe the employed code. In
Section~\ref{sec:results}  we present  our  calculations studying  the
dependence  of  the  evolution  of binary  systems  with  $(M_{NS})_i$
(Subsection~\ref{subsect:cbs-mns}) and discuss them in connection with
the    $P~-~M_{WD}$   relation    (Subsection~\ref{subsec:M-P}).    In
Section~\ref{sec:aplication}  we discuss  the  possibility of  finding
different initial  binary configurations  to account for  the observed
characteristics of systems containing a recycled pulsar and a low mass
WD  and,  as  an example,  we  study  the  case of  PSR~J0437-4715  in
detail.  Finally,  in  Section~\ref{sec:discu}  we  present  the  main
conclusions of this work.

\section{The computer code} \label{sec:numerical}

The code employed  here has been described elsewhere  (Benvenuto \& De
Vito  2003).  Briefly,  we  use a  generalized  Henyey technique  that
allows for the computation of  the stellar structure and mass transfer
episodes in a fully implicit  way. The code has an updated description
of  opacities, equation  of  state, nuclear  reactions and  diffusion,
while we simultaneously compute orbital evolution considering the main
processes of  angular momentum loss: angular momentum  carried away by
the matter lost from the system, gravitational radiation, and magnetic
braking.

Regarding the  inclusion of element diffusion, it  has several effects
on  the chemical  profile of  these stars,  especially in  the  WD and
pre~-~WD stages (see, e.g., Iben \& MacDonald 1985; Althaus, Serenelli
\&  Benvenuto 2001).  For example,  diffusion is  responsible  for the
occurrence of  almost pure  hydrogen atmospheres in  the case  of cool
enough DA~WDs.  Moreover, diffusion leads  to the sink of  hydrogen to
layers   hot  enough   for  triggering   the  occurrence   of  nuclear
burning. While in calculations neglecting diffusion, stellar models in
the  here  considered  mass   range  suffer  from  the  occurrence  of
(envelope)  hydrogen thermonuclear  flashes,  it has  been shown  that
diffusion forces the star to undergo supplementary flashes (Althaus et
al. 2001).

In our treatment  of the orbital evolution of  the system, we consider
that  the NS  is able  to retain  a fraction  $\beta$ of  the material
coming  from the  donor  star: $\dot{M}_{NS}=  -\beta \dot{M}$  (where
$\dot{M}$ is the  mass transfer rate from the donor  star), as done in
Benvenuto  \&  De  Vito  (2005).   We  consider  $\beta$  as  constant
throughout all RLOF episodes;  in particular, if not stated otherwise,
we set  $\beta =  0.5$, as  done in Podsiadlowski  et al.   (2002). We
assume  that material  lost from  the binary  system carries  away the
specific  angular momentum of  the compact  object ($\alpha=  1$; see,
e.g., Benvenuto \& De Vito 2003).

In  this work we  consider the  Mixing Length  Theory as  described in
Kippenhahn, Weigert  \& Hofmeister  (1967), setting the  Mixing Length
parameter to $l/H_P= 1.7432$  and including convective overshoot as in
Demarque et al.   (2004).   Furthermore,  we   consider  grey
atmospheres and  neglect the effects  of the irradiation of  the donor
star by the pulsar.

\section{Numerical results} \label{sec:results}

We select initial values for the system parameters (initial masses and
orbital  period) in  order  to obtain  systems  with HeWD  companions,
although some  of them evolve  to ultra-compact binaries  avoiding the
formation of  WDs.  The initial donor  star masses are  of 0.50, 0.65,
0.80, 1.00, 1.25, 1.50, 1.75,  2.00, 2.25, 2.50, 3.00, and 3.50~\msol,
of solar composition. We combine  these masses with accreting NSs with
initial masses $(M_{NS})_i$ of 0.8, 1.0, 1.2, and 1.4~\msol.

The  initial orbital  period for  the  three smaller  donor stars  are
of\footnote{This choice  is due to  the fact that, if  initial periods
were shorter, the Roche lobe would be smaller than the star even for a
Zero Age Main Sequence object; if they were longer, the star would not
fill the Roche lobe on the Hubble time.} 0.175, 0.20, and 0.30~d.  For
the other donor stars masses, initial periods are of 0.50, 0.75, 1.00,
1.50, 3.00, 6.00 and 12~d.  In all cases, the initial periods refer to
its value at the onset of  the first RLOF. Calculations start from the
Zero Age Main Sequence (we set  zero age there) and are followed up to
the formation of  a HeWD or an ultra-compact  system.  We computed the
evolution of the donor star up to  an age far in excess of Hubble time
of 20~Gyr,  or when the  donor has a  luminosity lower than  $1 \times
10^{-5}  L_{\odot}$. However in  some cases  we stop  the computations
earlier.  We do so if helium is ignited at the stellar core or if mass
transfer  becomes very intense  ($\dot{M} \geq  10^{-4}$~\msol/yr). In
Table~\ref{table:results}   we  present  the   main  results   of  our
calculations.

If   a   system   suffers   from   a   very   large   ($\dot{M}   \geq
10^{-4}$~\msol/yr), and still growing, mass transfer rate, we indicate
it in  this Table~\ref{table:results} as ``$\dot  M$ divergent''. This
behaviour  can be explained  in terms  of the  occurrence of  a common
envelope  (CE)  phase.  A CE  episode  can  be  the consequence  of  a
dynamical mass  transfer event. Dynamical mass  transfer is associated
typically with mass being transferred from the more massive component,
in a stage in which it  possesses a deep convective envelope (e.g., if
the onset  of a RLOF occurs  when the donor  star is on the  red giant
branch [RGB]  phase) or if the mass  ratio of the system  is large. In
such  conditions, the star  is unable  to contract  as rapidly  as its
Roche  lobe (in  fact  it  expands), thus  an  unstable mass  transfer
process ensues (Paczy\'nski \&  Sienkiewicz 1972). As a consequence of
the  high  accretion  rate,  the  accretor star,  driven  out  thermal
equilibrium, starts expanding (specially if the accretion rate exceeds
the Eddington limit) and fills  its own Roche lobe. The resulting mass
flow  leads to  the  formation  of the  CE  configuration (see,  e.g.,
Yungelson 1973, Webbink  1977, Livio 1989, Han \&  Webbink 1999). This
is the  case we find in  our calculations. The donor  star fills its
Roche lobe when is in the  RGB phase, with a deep convective envelope,
being donor star  the more massive component, and  with super Eddington
values  for $\dot  M$.  Then, we  consider  that this  leads  to a  CE
situation.  Also, divergent  $\dot M$ episodes are found  for the case
of donor status  with very short orbital periods  and masses $M_i \geq
3.0$~\msol. For  these systems,  the onset of  the RLOF  occurs during
core  hydrogen  burning and  should  be  associated  with a  ``delayed
dynamical''  unstable mass transfer  as found  by Podsiadlowski  et al
(2002).  Notice  that ``$\dot  M$  divergent''  systems  are found  at
shorter  initial orbital  periods the  lighter is  the NS.  This again
indicates  that the  evolution of  the systems  heavily depend  on the
initial mass of the NS.

\onecolumn
\begin{landscape}
\begin{center}
\begin{longtable}{|c|r|rcc|rcc|rcc|rcc|}
\caption{\label{table:results}  Main results  of our  binary evolution
calculations.  First and  second columns list the initial  mass of the
donor   star  and   the  initial   orbital  period   of   the  systems
respectively. For  each donor star  and orbital period we  compute the
evolution of  binary systems  for different values  of the  initial NS
mass:  $(M_{NS})_i=$ 0.8, 1.0,  1.2, and  1.4~\msol.  For  each system
that evolves to  ultra~-~compact or HeWD~-~MSP pair we  list the final
period, the  donor remnant and  NS masses.  Numbers in  italics denote
systems   that    form   a   WD    but   we   do   not    include   in
Fig.~(\ref{fig:mass-porb}).  For further details see the main text.}\\
\hline
 \multicolumn{1}{|c|}{} & 
 \multicolumn{1}{c|}{}  & 
 \multicolumn{3}{c|}{$(M_{NS})_i = 0.80 M_{\odot}$} & 
 \multicolumn{3}{c|}{$(M_{NS})_i = 1.00 M_{\odot}$} &
 \multicolumn{3}{c|}{$(M_{NS})_i = 1.20 M_{\odot}$} & 
 \multicolumn{3}{c|}{$(M_{NS})_i = 1.40 M_{\odot}$} \\
\hline
\hline
$M_i$ & $P_{i}$ & $P_{f}$ & $M_{WD}$ & $M_{NS}$ & $P_{f}$ & $M_{WD}$ & $M_{NS}$ & $P_{f}$ & $M_{WD}$ & $M_{NS}$ & 
$P_{f}$ & $M_{WD}$ & $M_{NS}$ \\
$[M_{\odot}]$ & $[d]$ & $[d]$ & $[M_{\odot}]$ & $[M_{\odot}]$ & $[d]$ & $[M_{\odot}]$ & $[M_{\odot}]$ & $[d]$ &
$[M_{\odot}]$ & $[M_{\odot}]$ & $[d]$ & $[M_{\odot}]$ & $[M_{\odot}]$ \\
\hline
\hline
\endfirsthead

\caption{Continued}\\

\hline
 \multicolumn{1}{|c|}{} & 
 \multicolumn{1}{c|}{}  & 
 \multicolumn{3}{c|}{$(M_{NS})_i = 0.80 M_{\odot}$} & 
 \multicolumn{3}{c|}{$(M_{NS})_i = 1.00 M_{\odot}$} &
 \multicolumn{3}{c|}{$(M_{NS})_i = 1.20 M_{\odot}$} & 
 \multicolumn{3}{c|}{$(M_{NS})_i = 1.40 M_{\odot}$} \\
\hline
\hline
$M_i$ & $P_{i}$ & $P_{f}$ & $M_{WD}$ & $M_{NS}$ & $P_{f}$ & $M_{WD}$ & $M_{NS}$ & $P_{f}$ & $M_{WD}$ & $M_{NS}$ & 
$P_{f}$ & $M_{WD}$ & $M_{NS}$ \\
$[M_{\odot}]$ & $[d]$ & $[d]$ & $[M_{\odot}]$ & $[M_{\odot}]$ & $[d]$ & $[M_{\odot}]$ & $[M_{\odot}]$ & $[d]$ &
$[M_{\odot}]$ & $[M_{\odot}]$ & $[d]$ & $[M_{\odot}]$ & $[M_{\odot}]$ \\
\hline
\hline
\endhead

\hline
\multicolumn{14}{r}{(Continue in the next page)}\\
\endfoot

\hline
\endlastfoot
0.50 &0.175 &  0.0585& 0.0366 & 1.0317 & 0.0595& 0.0362 & 1.231 &  0.0603& 0.0355 & 1.4322  &  0.0611& 0.0351 & 1.6325 \\ 
\hline
0.65 & 0.20 &  0.0589& 0.0368 & 1.1066 & 0.0602& 0.0362 & 1.306 &  0.0609& 0.0356 & 1.5072  &  0.0616& 0.0351 & 1.7075 \\ 
\hline 
0.80&  0.30 &  0.0537& 0.0323 & 1.1839 & 0.0568& 0.0325 & 1.383 &  0.0571& 0.0315 & 1.5842  &  0.0485& 0.0447 & 1.7777 \\ 
\hline 
     & 0.50 &  0.0274& 0.0305 & 1.2848 & 0.0361& 0.0154 & 1.492 &  0.0384& 0.0154 & 1.6923  &  0.0365& 0.0148 & 1.8926 \\ 
     & 0.75 &  1.8603& 0.1878 & 1.2053 & 2.2487& 0.1918 & 1.403 &  2.5233& 0.1966 & 1.6010  &  2.9950& 0.2011 & 1.7983 \\
     & 1.00 &  4.7508& 0.2180 & 1.1755 & 6.0133& 0.2258 & 1.386 &  6.9490& 0.2287 & 1.5829  &  8.1373& 0.2317 & 1.7831 \\
1.00 & 1.50 &  9.5741& 0.2360 & 1.1251 &12.6974& 0.2420 & 1.371 & 15.5468& 0.2471 & 1.5673  & 17.7987& 0.2505 & 1.7706 \\
     & 3.00 & 21.6577& 0.2593 & 1.0832 &29.3771& 0.2667 & 1.349 & 35.0054& 0.2719 & 1.5630  & 39.8687& 0.2759 & 1.7611 \\
     & 6.00 & 40.9639& 0.2778 & 1.0471 &55.4442& 0.2871 & 1.323 & 66.4038& 0.2929 & 1.5486  & 75.4193& 0.2972 & 1.7506 \\
     &12.00 & 73.3650& 0.2971 & 1.0098 &98.2015& 0.3079 & 1.283 &117.7680& 0.3141 & 1.5229  &133.8719& 0.3187 & 1.7270 \\ 
\hline
     & 0.50 &  0.0445& 0.0408 & 1.4046 & 0.0526& 0.0286 & 1.610 &  0.0532& 0.0277 & 1.8112  &  0.0546& 0.0281 & 2.0109 \\ 
     & 0.75 & {\it 0.0407} & {\it 0.1414}  & {\it 1.3543}  & {\it 0.0516 } & {\it 0.1531 } & {\it 1.548 } &  {\it 0.0336 } & {\it 0.1609 } & {\it 1.7445 } &  0.4501& 0.1672 & 1.9403 \\
     & 1.00 &  4.4800& 0.2172 & 1.2943 & 5.4725& 0.2229 & 1.512 &  6.0909& 0.2253 & 1.7115  &  6.6807& 0.2273 & 1.9106 \\
1.25 & 1.50 &\multicolumn{3}{c|}{$\dot M$ divergent} &15.8528& 0.2472 & 1.413 & 18.9614& 0.2546 & 1.6786  & 22.2105& 0.2587 & 1.8931 \\
     & 3.00 &        &        &        &\multicolumn{3}{c|}{$\dot M$ divergent}& 44.2353& 0.2789 & 1.6365 & 52.2558& 0.2841 & 1.8746 \\
     & 6.00 &        &        &        &       &	&       & 81.7321& 0.3003 & 1.5641  & 97.9597& 0.3062 & 1.8424 \\
     &12.00 &        &        &        &       &	&       &144.7212& 0.3223 & 1.5527  &169.8661& 0.3277 & 1.7152 \\ 
\hline
     & 0.50 &  0.0590& 0.0335 & 1.5332 & 0.0475& 0.0434 & 1.728 &  0.0465& 0.0389 & 1.9305  &  0.0545& 0.0278 & 2.1361 \\ 
     & 0.75 &  0.0483& 0.0245 & 1.5377 & 0.0466& 0.0220 & 1.739 &  0.0475& 0.0199 & 1.9401  &  0.0496& 0.0197 & 2.1401 \\
     & 1.00 &  {\it 0.0403}  & {\it 0.1601}  & {\it 1.4489}  & 0.5110& 0.1741 & 1.660 &  1.6702& 0.1943 & 1.8521  &  4.0510& 0.2047 & 2.0468 \\
1.50 & 1.50 &\multicolumn{3}{c|}{$\dot M$ divergent}&\multicolumn{3}{c|}{$\dot M$ divergent}& 28.9438& 0.2693 & 1.7149 
                                                                                            & 31.7147& 0.2708 & 2.0139 \\
     & 3.00 &        &        &        &       &	&       & 52.5323& 0.2884 & 1.5327  & 65.1049& 0.2941 & 1.8746 \\
     & 6.00 &        &        &        &       &	&       &\multicolumn{3}{c|}{$\dot M$ divergent}&111.8036& 0.3134 & 1.8673 \\
     &12.00 &        &        &        &       &	&       &	 &	  &	    &198.2880& 0.3367 & 1.7854 \\
\hline
     & 0.50 &  0.0588& 0.0337 & 1.4818 & 0.0590& 0.0321 & 1.782 &  0.0468& 0.0389 & 2.0357  &  0.0530& 0.0262 & 2.2619 \\ 
     & 0.75 &  0.0501& 0.0269 & 1.4823 & 0.0494& 0.0246 & 1.785 &  0.0473& 0.0211 & 2.0426  &  0.0493& 0.0197 & 2.2652 \\
     & 1.00 &  0.0460& 0.0219 & 1.5051 & 0.0465& 0.0205 & 1.820 &  0.0491& 0.0186 & 2.0657  &  {\it 0.0490 }& {\it 0.1648 } & {\it 2.1913 } \\
1.75 & 1.50 & 10.5640& 0.2459 & 1.1158 &22.5014& 0.2537 & 1.500 & 34.1985& 0.2630 & 1.8120  & 40.5855& 0.2814 & 2.0920 \\
     & 3.00 &\multicolumn{3}{c|}{$\dot M$ divergent}&\multicolumn{3}{c|}{$\dot M$ divergent}&\multicolumn{3}{c|}{$\dot M$ divergent} 
                                                                                            & 68.7946& 0.3104 & 1.6614 \\
     & 6.00 &        &        &        &       &	&       &	 &	  &	    &\multicolumn{3}{c|}{$\dot M$ divergent}\\
\hline
     & 0.50 &  0.0583& 0.0345 & 1.2849 & 0.0603& 0.0332 & 1.732 &  0.0564& 0.0288 & 2.0303  &  0.0523& 0.0258 & 2.3501 \\
     & 0.75 &  0.0504& 0.0279 & 1.3204 & 0.0518& 0.0273 & 1.714 &  0.0481& 0.0228 & 2.0155  &  0.0487& 0.0196 & 2.3257 \\
2.00 & 1.00 &  0.0464& 0.0233 & 1.3209 & 0.0472& 0.0227 & 1.717 &  0.0482& 0.0196 & 2.0160  &  0.0324& 0.0268 & 2.2870 \\
     & 1.50 &  0.0301& 0.0266 & 1.3108 & 1.1441& 0.2047 & 1.670 &  3.6701& 0.2317 & 1.9679  & 17.7005& 0.2621 & 2.2028 \\
     & 3.00 &\multicolumn{3}{c|}{$\dot M$ divergent}& \multicolumn{3}{c|}{$\dot M$ divergent} & \multicolumn{3}{c|}{$\dot M$ divergent} 
                                                                              & \multicolumn{3}{c|}{$\dot M$ divergent}\\
\hline
     & 0.50 &\multicolumn{3}{c|}{$\dot M$ divergent}& 0.0602& 0.0343 & 1.644 &  0.0594& 0.0315 & 1.9949  &  0.0535& 0.0264 & 2.3217 \\ 
     & 0.75 &        &        &        & 0.0523& 0.0282 & 1.637 &  0.0500& 0.0248 & 1.9593  &  0.0472& 0.0201 & 2.2686 \\
2.25 & 1.00 &\multicolumn{3}{c|}{in  all}& 0.0476& 0.0236 & 1.610 &  0.0477& 0.0219 & 1.9453  &  0.0517& 0.0196 & 2.2417 \\
     & 1.50 &        &        &        & 0.0265& 0.0224 & 1.589 &  2.6496& 0.2176 & 1.8582  &  6.6082& 0.2427 & 2.1434 \\
     & 3.00 &\multicolumn{3}{c|}{cases}&\multicolumn{3}{c|}{He burning}&\multicolumn{3}{c|}{He burning}&\multicolumn{3}{c|}{He  burning}\\
\hline  
     & 0.50 &\multicolumn{3}{c|}{$\dot M$ divergent}& 0.0604& 0.0348 & 1.488 &  0.0614& 0.0338 & 1.9351  &  0.0468& 0.0378 & 2.2212 \\
     & 0.75 &        &        &        & 0.0526& 0.0289 & 1.524 &  0.0416& 0.0412 & 1.8920  &  0.0479& 0.0209 & 2.2090 \\
2.50 & 1.00 &\multicolumn{3}{c|}{in  all}& 0.0477& 0.0246 & 1.505 &  0.0486& 0.0234 & 1.8692  &  0.0487& 0.0196 & 2.1790 \\
     & 1.50 &        &        &        & 2.7871& 0.2170 & 1.355 &  3.6504& 0.2217 & 1.7351  &  6.1671& 0.2367 & 2.0551 \\
     & 3.00 &\multicolumn{3}{c|}{cases}&\multicolumn{3}{c|}{He burning}&\multicolumn{3}{c|}{He burning}&\multicolumn{3}{c|}{He burning}\\ 
\hline
     & 0.50 &\multicolumn{3}{c|}{$\dot M$ divergent}&\multicolumn{3}{c|}{$\dot M$ divergent}&\multicolumn{3}{c|}{$\dot M$ divergent}
                                                                                              &  0.0611& 0.0327 & 2.1107 \\
     & 0.75 &        &        &        &       &	&       &  0.0491& 0.0242 & 1.7349  &  0.0490& 0.0229 & 2.1056 \\
3.00 & 1.00 &\multicolumn{3}{c|}{in  all}&\multicolumn{3}{c|}{in  all}&  0.0468& 0.0184 & 1.7217  &  4.5711& 0.2053 & 1.9641 \\
     & 1.50 &        &        &        &       &	&       & 12.6423& 0.2572 & 1.5379  & 15.4657& 0.2612 & 1.8645 \\
     & 3.00 &\multicolumn{3}{c|}{cases}&\multicolumn{3}{c|}{cases}&\multicolumn{3}{c|}{He burning}&\multicolumn{3}{c|}{He burning}\\ 
\hline
     & 0.50 &\multicolumn{3}{c|}{$\dot M$ divergent}&\multicolumn{3}{c|}{$\dot M$ divergent}&\multicolumn{3}{c|}{$\dot M$ divergent}
                                                                                           & \multicolumn{3}{c|}{$\dot M$ divergent}\\
     & 0.75 &        &        &        &       &	&       &	 &	  &	    &  0.0479& 0.0197 & 1.9640 \\
3.50 & 1.00 &\multicolumn{3}{c|}{in  all}&\multicolumn{3}{c|}{in  all}&\multicolumn{3}{c|}{in  all}& 15.9734& 0.2378 & 1.8293 \\
     & 1.50 &        &        &        &       &	&       &	 &	  &	    & 21.0553& 0.2755 & 1.7556 \\
     & 3.00 &\multicolumn{3}{c|}{cases}&\multicolumn{3}{c|}{cases}&\multicolumn{3}{c|}{cases}&\multicolumn{3}{c|}{He burning}\\ 
\hline
\end{longtable}
\end{center}
\end{landscape}
\twocolumn

\subsection{The dependence of the evolution of close binary systems upon the
initial mass of the neutron star}\label{subsect:cbs-mns}

\begin{figure}   
\epsfysize=400pt  
\epsfbox{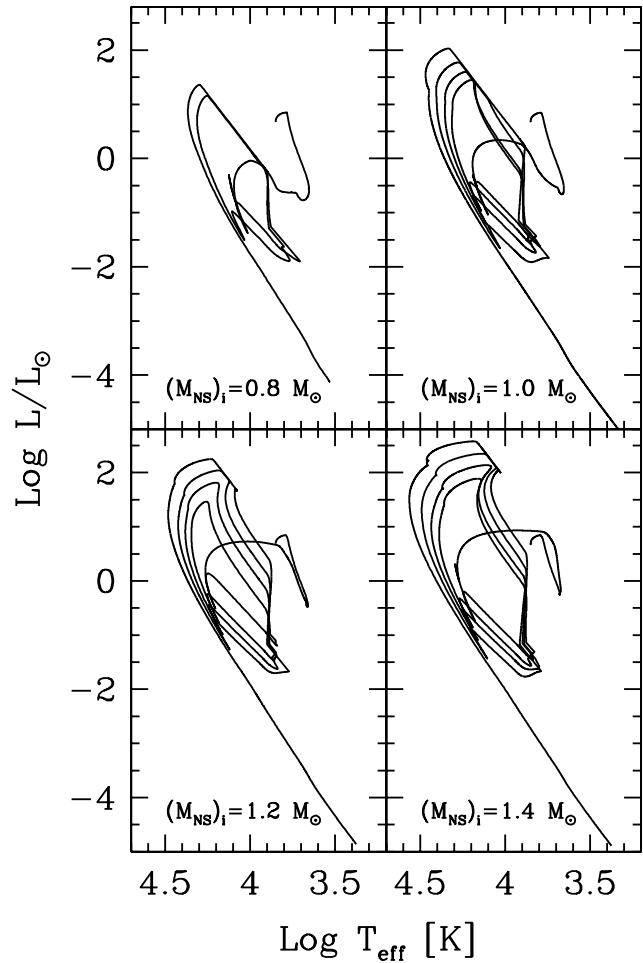} 
\caption{\label{fig:hr_15_10}  The evolutionary  tracks  for a  normal
donor star with  initial mass of 1.5~\msun evolving  in binary systems
with different  initial NS  masses. The initial  orbital period  is of
1~day. The loops  are due to hydrogen thermonuclear  flashes (see main
text for further details).}
\end{figure}

Among  the  results  presented  in Table~\ref{table:results},  we  may
select  a subset  of evolutionary  calculations, for  a  given initial
donor star mass and orbital period to study the behaviour of CBSs when
we change the initial NS mass.  In Table~\ref{table:data_for_15_10} we
present supplementary data for the  case of a donor star of 1.5~\msol,
initial  period of  1~day  and  different values  for  the initial  NS
mass.  In Fig.~(\ref{fig:hr_15_10})  we show  the  evolutionary tracks
corresponding     to     the     binary    systems     included     in
Table~\ref{table:data_for_15_10}.  In all of  the selected  cases, the
donor  star undergoes several  thermonuclear hydrogen  flashes.  These
flashes are  the responsible for  the quasi~-~cyclic behaviour  in the
Hertzsprung~-~Russell diagram. Let us  briefly quote that these events
are due to  the heating of the bottom of the  hydrogen envelope of the
(then) pre~-~WD  object. At that  place matter is  degenerate, forcing
the  onset of  unstable nuclear  burning. For  further details  on the
evolution  of a  pre~-~WD object  undergoing thermonuclear  flashes in
presence of diffusion, see Althaus et al.  (2001).

\begin{centering} 
\begin{table*} 
\caption{\label{table:data_for_15_10}  Main   characteristics  of  the
evolution  of  systems  that  initially  have a  donor  star  mass  of
1.5~\msol, an  orbital period  of 1~day and  different values  for the
initial NS  mass. From left to right  we list the initial  mass of the
accreting NS, the time for the onset of the first RLOF, the time spent
during this  RLOF, the final values of  the WD and the  NS masses, and
the final orbital period of the system.}
\begin{tabular}{cccccc}
\hline 
\hline
$(M_{NS})_i$   & $t_{i-\dot M}$ & $\Delta t_{\dot M}$ & $M_{WD}$      & $M_{NS}$      & $P$ \\
$[M_{\odot}]$ & $[Gyr]$        & $[Gyr]$             & $[M_{\odot}]$ & $[M_{\odot}]$ & $[d]$ \\
\hline
0.80 & 2.740 & 2.295 & 0.1601 & 1.4489 & 0.0403 \\
1.00 & 2.624 & 1.886 & 0.1741 & 1.6608 & 0.5110 \\
1.20 & 2.595 & 1.371 & 0.1943 & 1.8521 & 1.6702 \\
1.40 & 2.529 & 1.055 & 0.2047 & 2.0468 & 4.0510 \\
\hline 
\hline 
\end{tabular} 
\end{table*} 
\end{centering}

\begin{figure}  
\epsfysize=300pt  
\epsfbox{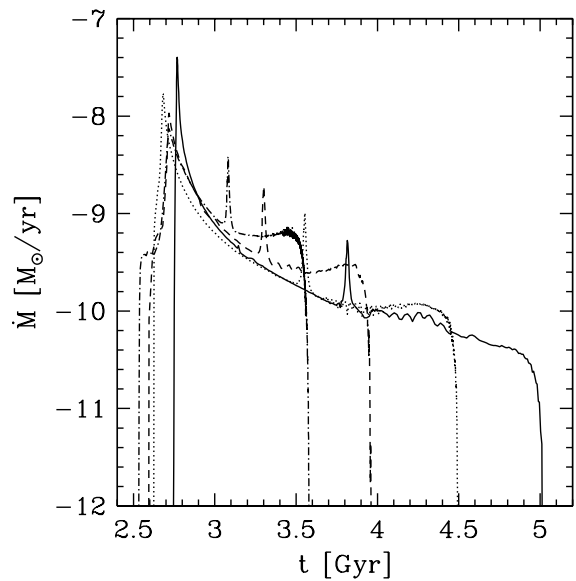}  
\caption{\label{fig:mdot_15_10}  The temporal  evolution  of the  mass
transfer rate for the systems considered in Fig.~(\ref{fig:hr_15_10}).
Solid, dot,  short~-~dash, and dot~-~short~-~dash lines  show the mass
transfer rates  of the objects  corresponding to initial NS  masses of
0.8, 1.0, 1.2, and 1.4~\msun  respectively. Here we show only the mass
loss episodes not induced by  thermonuclear flashes in the envelope of
the star.}
\end{figure}

\begin{figure}  
\epsfysize=300pt  
\epsfbox{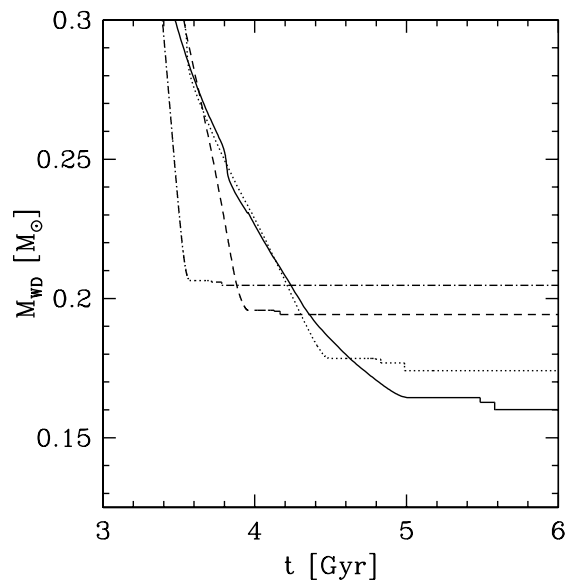}  
\caption{\label{fig:mwd_15_10} Evolution of the mass of the donor star
with  time, for  systems depicted  in  Fig.~(\ref{fig:hr_15_10}).  The
meaning of lines are the same of in Fig.~(\ref{fig:mdot_15_10}).}
\end{figure} 

\begin{figure}  
\epsfysize=300pt  
\epsfbox{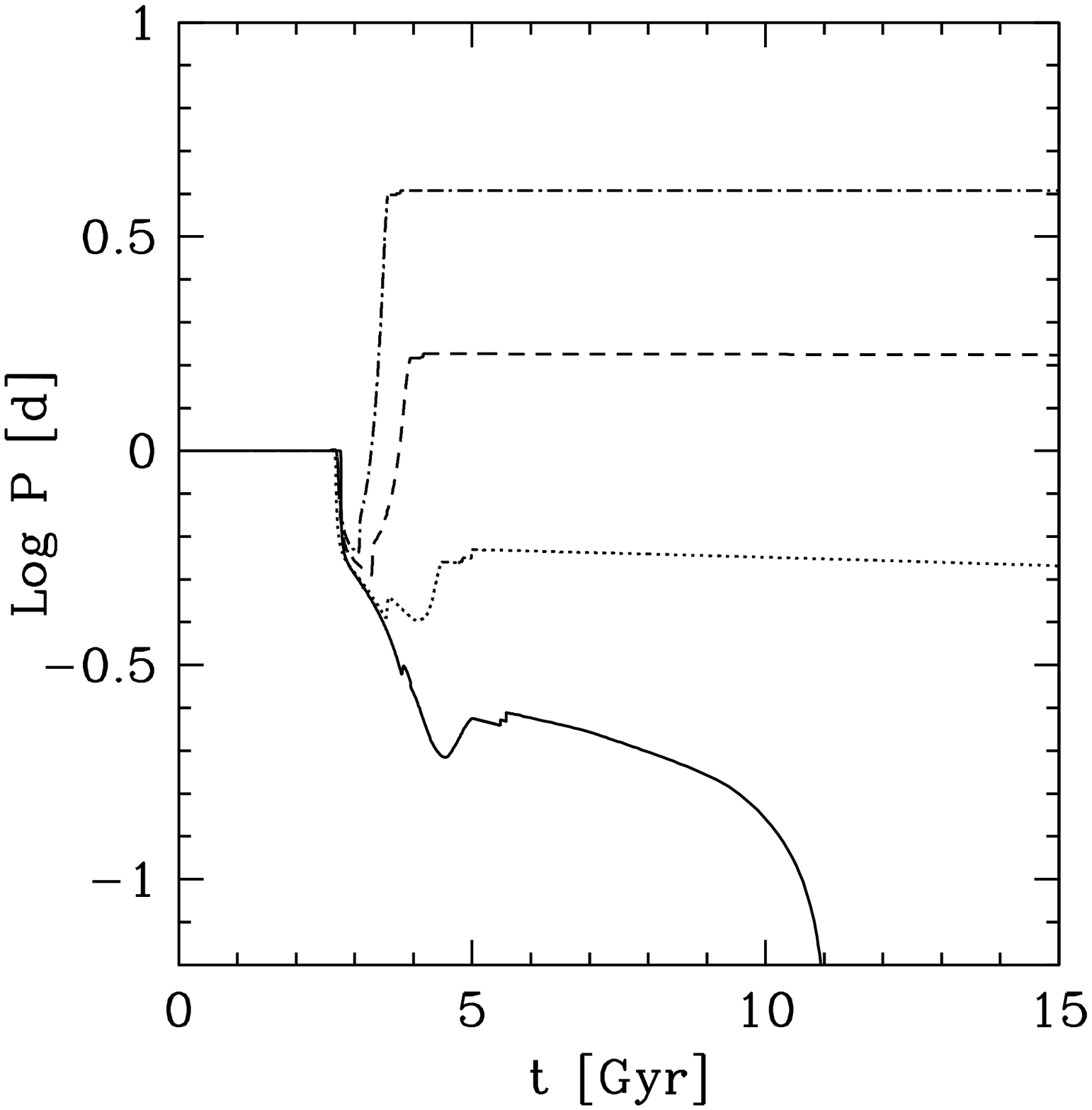} 
\caption{\label{fig:pt_15_10} The evolution  of the orbital period for
the  systems  depicted in  Fig.~(\ref{fig:hr_15_10}).  The meaning  of
lines are the same of in Fig.~(\ref{fig:mdot_15_10}).}
\end{figure}

In Fig.~(\ref{fig:mdot_15_10}) we show the mass loss rate for the same
set of  systems. The lower is  the initial mass for  the accreting NS,
the longer is the time spent by the system in the RLOF episode. Notice
that the  onset of the RLOF occurs  later the less massive  is the NS,
because  the  Roche lobe  of  the donor  star  is  bigger (see,  e.g.,
Eggleton  1983). We  find less  massive WD  remnants for  less massive
accreting   NS,   as  we   can   see  in   Fig.~(\ref{fig:mwd_15_10}).
Fig.~(\ref{fig:pt_15_10}) shows the evolution of the orbital period as
a  function of  time for  the  same subset  of systems.   We see  that
systems having less massive NSs evolve to tighter configurations.

From the  results presented  above we find  that the evolution  of the
donor  star {\it  heavily} depends  on the  value of  the mass  of the
NS. This is one of the main findings of the present paper.

\subsection{The orbital period~-~WD mass relation} \label{subsec:M-P}

As stated  above, one of the aims  of the present paper  is to explore
the dependence of the $P~-~M_{WD}$  relation upon the initial NS mass.
Rappaport et  al.  (1995)  claim that this  relation should  be fairly
insensitive to changes in the initial NS mass. Their argument is based
on  the well known  fact that  there exist  a somewhat  tight relation
between  the mass and  the radius  of the  cores of  low~-~mass giants
(see,  e.g.,  Joss,  Rappaport   \&  Lewis  1987).  Clearly,  this  is
applicable only for the case of  donor stars that undergo the onset of
the  RLOF  as  giants.  Consequently,  our  calculations  provide  the
opportunity to  test the validity  of the conclusions of  Rappaport et
al.  (1995) in a quantitative way, at least for the case of HeWDs.

Let us repeat, for the sake of completeness, the argument of Rappaport
et al.   (1995) in detail. For  the kind of binary  systems studied in
this  work,  the  orbit   is  considered  circular  because  of  tidal
dissipation since the onset of the first RLOF and should remain nearly
circular thereafter.   During later phases of mass  transfer (once the
mass of  the donor star  has become smaller  than that of the  NS), an
approximate expression  for the  radius of the  Roche lobe,  $R_L$, in
terms of the constituent masses is given by (Paczy\'nski 1971)

\begin{equation} \label{eq:rochelobe}
R_L = 0.46 a \Big(1 + {{M_{NS}}\over{M_G}}\Big)^{-1/3},
\end{equation}

\noindent where  $M_G$ the mass of  the giant, and $a$  is the orbital
separation.   If  we  combine Eq.~(\ref{eq:rochelobe})  with  Kepler's
third law and set $R_G = R_L$ (i.e. the giant fills its Roche lobe) we
obtain an expression for the orbital period

\begin{equation} \label{eq:porb}
P= 20 \, G^{-1/2} \, R_G^{3/2} \, M_G^{-1/2}.
\end{equation}

\noindent Note that $P$ is independent of the mass of the NS. Near the
end of  the mass  transfer phase,  the envelope of  the giant  is very
tenuous and  embraces a  mass substantially smaller  than that  of the
core, $M_c$; then $M_G \simeq M_c$. Therefore, since $R_G$ is a nearly
unique function of $M_c$, the  final orbital period at the termination
of the mass transfer can be written as

\begin{equation} \label{eq:pmass}
P \simeq 20 \, G^{-1/2} \, R_G^{3/2}(M_c) \, M_{c}^{-1/2}
\end{equation} 

In order  to test the  relation given by  Eq.~(\ref{eq:pmass}) against
the  observed set of  binary pulsars  containing low  mass WDs,  it is
important  to  establish an  accurate  theoretical core  mass~-~radius
relation  ($M_c -  R_G$).  Rappaport  et al.  (1995) have  performed a
systematic study of  the core mass~-~radius relation from  a series of
single star evolutionary calculations.   They covered a range of giant
masses   between   0.8    and   2.0~\msun   and   different   chemical
compositions. They fitted the relation  $M_c - R_G$ with the empirical
formula

\begin{equation} \label{eq:mcore-radio}
R_G(m_c) = \left[ R_0 \, {{m_c^{4.5}}\over{1 +  m_c^4}} + 0.5 \right] 
           \, R_{\odot}
\end{equation}

\noindent where  $m_c =  M_c / M_{\odot}$  and $R_0$ is  an adjustable
constant that for the case of  Population I objects takes the value of
to $R_0  = 5500$.  Now,  by combining the core  mass~-~radius relation
(Eq.~\ref{eq:mcore-radio}) and  (Eq.~\ref{eq:pmass}), and setting $M_c
= M_{WD}$, $M_{WD}$ being the mass of the WD that has been the core of
the giant before envelope dissipation, we obtain

\begin{equation} \label{eq:mass-p-relation}
P \simeq 0.374 \left[R_0 \, {{m_{WD}^{4.5}}\over{1 +  m_{WD}^4}} + 0.5 \right]^{3/2}
             m_{WD}^{-1/2} \, d
\end{equation}

\noindent where $m_{WD} = M_{WD} / M_{\odot}$.

Since then,  other relations have  been presented. Tauris  \& Savonije
(1999) given the relation

\begin{equation} \label{eq:mass-p-relation-tauris}
{m_{WD}} = \left( {{P}\over{b}} \right)^{1/a} + c
\end{equation}

\noindent were,  for the case of  Population I, the authors  find $a =
4.50$, $b = 1.2 \times 10^5$, $c=0.120$, with $P$ and $b$ expressed in
days.   This fit  is valid  for $0.18  \leq m_{WD}  \leq  0.45$. Also,
Nelson, Dubeau \& MacCannell (2004) stated the relation

\begin{equation} \label{eq:mass-p-relation-nelson}
P= 0.1042 \, Z^{0.3} \, 10^{(10.7 \, m_{WD})} \, d,
\end{equation}

\noindent were  $Z$ is the metal  content of the donor  star. In this
case, the fit is valid for $m_{WD} \geq 0.25$.

\begin{figure*}   
\epsfysize=450pt  
\epsfbox{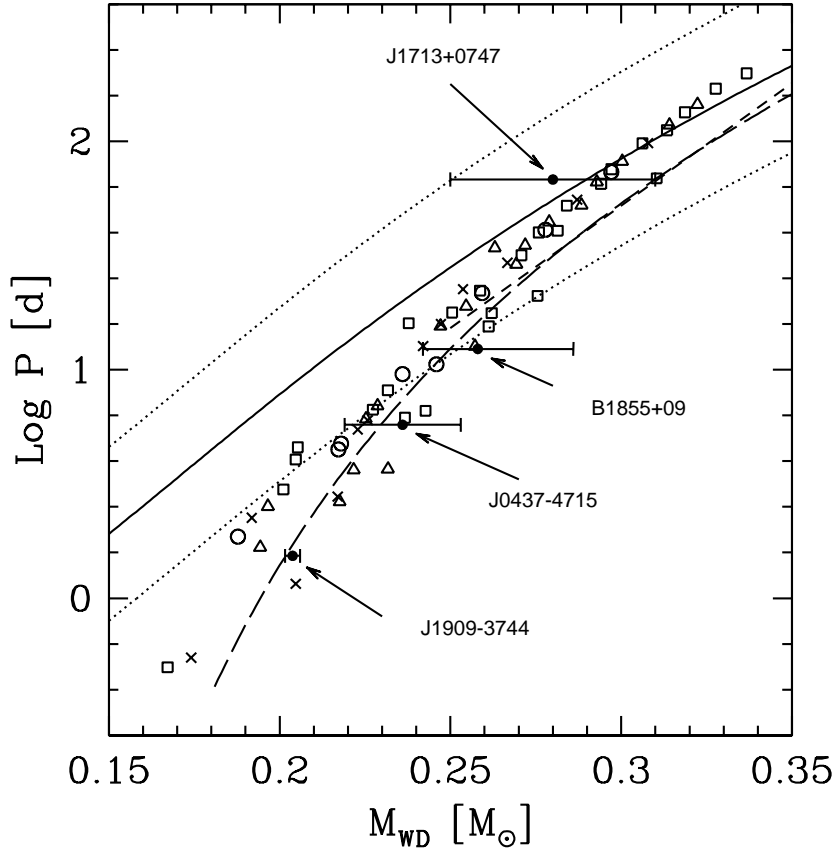} 
\caption{\label{fig:mass-porb}  The   $P~-~M_{WD}$  relation  for  the
binary systems presented in this work. Circles, crosses, triangles and
squares depict  systems with  accreting NS of  initial mass  of $0.80,
1.00, 1.20$ and $1.40$~\msol,  respectively. In addition, we plot with
solid  line  the  relation  given  by Rappaport  et  al.   (1995)  for
Population  I, with  their error  bars  (dot lines),  the relation  of
Tauris \& Savonije  (1999) with long~-~dash line, and  the relation of
Nelson, Dubeau  \& MacCannell (2004) with short~-~dash  line.  Also we
have   included  the  WDs   masses  and   orbital  periods   cited  in
Table~\ref{table:shapiro-delay-data}  with   the  corresponding  error
bars.}
\end{figure*}

In  Fig.~(\ref{fig:mass-porb}) we plot  the $P~-~M_{WD}$  relation for
some  of our  models, where  we also  include the  relations  given by
Eq.~(\ref{eq:mass-p-relation}) for Population  I with their error bars
given by Rappaport  et al. (1995) and also that  of Tauris \& Savonije
(1999) (Eq.~\ref{eq:mass-p-relation-tauris}) and  Nelson et al. (2004)
(Eq.~\ref{eq:mass-p-relation-nelson}).    As  it   can   be  seen   in
Fig.~(\ref{fig:mass-porb}),  our evolutionary calculations  agree with
the  prediction that in  wide binaries,  the $P~-~M_{WD}$  relation is
fairly independent to the value of the initial NS mass.

In  Fig.~(\ref{fig:mass-porb})  we  have  not  included  some  of  our
models. The criterion to include a model was simply if the position in
the $P~-~M_{WD}$ plane  is fairly independent of time  on a reasonably
large time interval. For example,  as stated above, some of our models
evolve to  ultra-compact systems with  masses of only few  percents of
the solar mass.  Even for the  dimmest considered models they are on a
RLOF episode, and thus move  on the aforementioned plane. In any case,
it  is clear that  these object  are quite  different from  those that
represent our main interest. Notably, there is another kind of objects
that do form a  HeWD but on a very tight orbit.  Data related to these
objects  is  presented in  Table~\ref{table:results}  with numbers  in
italics. These systems are subject  to strong orbital evolution due to
gravitational wave radiation. As they  are not on a RLOF episode, they
evolve  downwards  vertically.   Thus,  in studying  the  $P~-~M_{WD}$
relation we  shall consider systems with  a period $P  > 0.25$~d.  For
systems with $P < 1$~d we considered the value of $P$ at 13~Gyr, while
for the others this is an irrelevant detail.

\begin{figure}  
\epsfysize=300pt  
\epsfbox{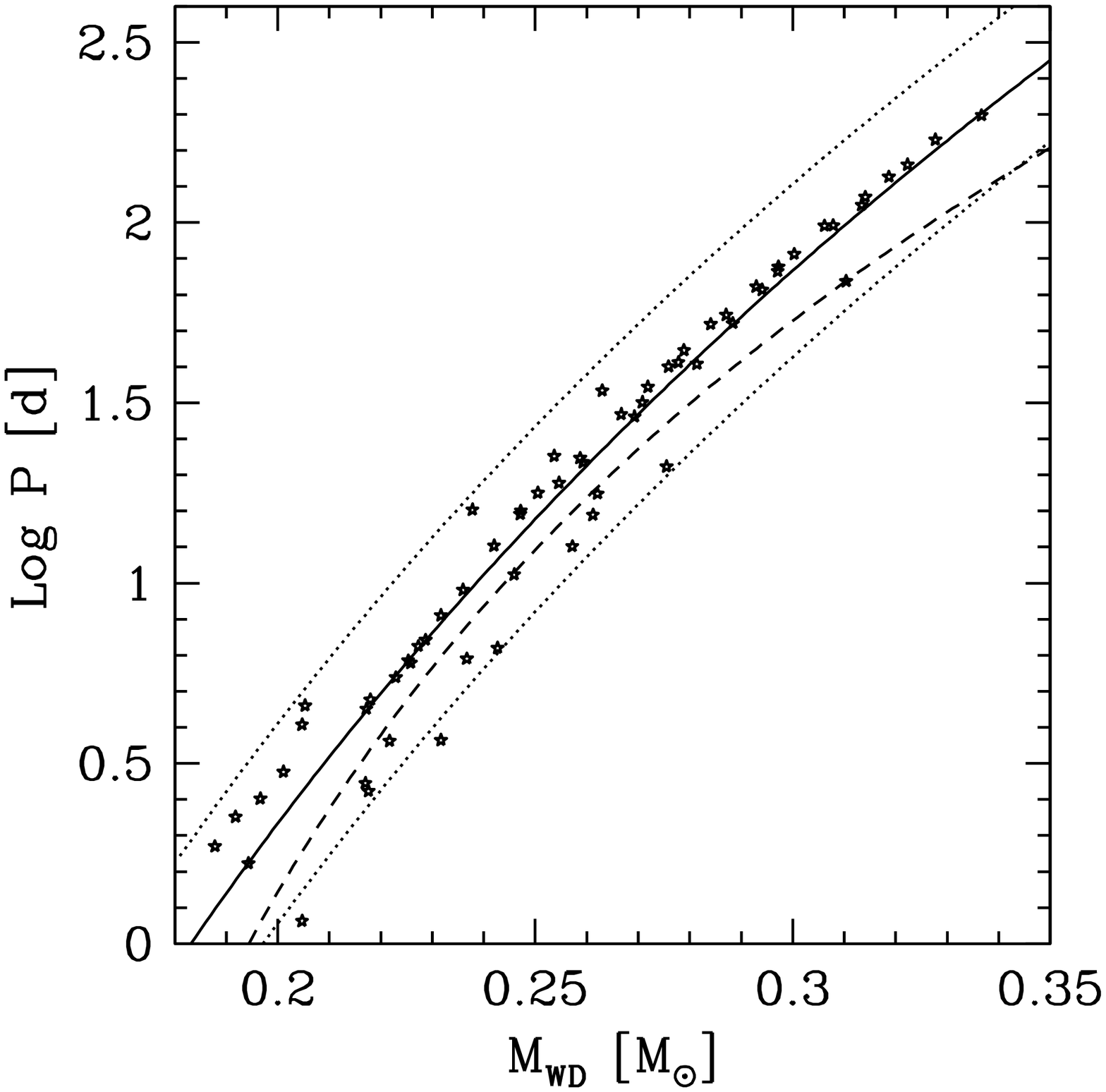} 
\caption{\label{fig:ajuste-nos-ts} The  fit of our  results, performed
with  a linear  function (Eq.~\ref{eq:nuestrofit})  in the  plane $Log
P~-~Log (M_{WD})$ denoted  with a solid line. Upper  and lower limits,
showing  the uncertainty  inherent to  our fit  are given  with dotted
lines. Short~-~dash line represents the $P~-~M_{WD}$ relation found by
Tauris \& Savonije (1999).}
\end{figure}

Now  we  shall  present a  fit  of  our  results in  the  $P~-~M_{WD}$
plane.  Notably,  the values  of  $\log{m_{WD}}$  have an  approximate
linear dependence  upon $\log{P}$.  Thus, we propose  a linear  fit by
least squares. The fit we find is

\begin{equation} \label{eq:nuestrofit}
P= B \; (m_{WD})^{A}\; d 
\end{equation}

\noindent  where   $A=  8.7078$,   $B=  2.6303  \times   10^{6}$,  see
Fig.~(\ref{fig:ajuste-nos-ts}).  In this Figure we also included, with
dotted lines,  the uncertainty associated with  this fit corresponding
to  1~$\sigma$ deviation for  the coefficients  $A$ and  $\log{B}$ for
which  $(A, B)=  (8.4948,  3.5372 \times  10^{6})$  (upper curve)  and
$(8.9208, 1.9559 \times 10^{6})$ (lower curve).  This relation is very
similar     to     that    of     Tauris     \&    Savonije     (1999)
(Eq.~\ref{eq:mass-p-relation-tauris}),  although  it  accommodates  to
periods slightly  longer. In any  case, the differences  between their
fit        (Eq.~\ref{eq:mass-p-relation-tauris})        and       ours
(Eq.~\ref{eq:nuestrofit}) are smaller than the uncertainty in our fit.
Thus, we consider that the agreement is fairly good.  On the contrary,
our fit (Eq.~\ref{eq:nuestrofit}) is noticeably different from the one
presented by Rappaport et al.   (1995). While for $m_{WD} \approx 0.3$
our calculations are in good agreement with their fit, this is not the
case for lower WD mass  values.  This result is not surprising, simply
because the  fit presented  by Tauris \&  Savonije (1999) is  based on
full  binary evolution calculations,  while that  of Rappaport  et al.
(1995) relies  on single  stellar evolution  results. Finally,  in the
range of masses $m_{WD} >  0.25$ the agreement between the relation of
Nelson et al.  (2004) (Eq.~\ref{eq:mass-p-relation-nelson}) and our is
also good although it is a bit  poorer than for the case of the others
analysed previously.

\section{Application to MSP - WD systems} \label{sec:aplication}

In previous works  (Benvenuto \& De Vito 2005;  Benvenuto et al. 2006)
we have tried to identify  possible binary system progenitors for some
of the best observed MSP~-~WD systems (PSR~J0437-4715, PSR~J1713+0747,
and PSR~B1855+09).  In  those papers we assumed a  canonical value for
the initial NS mass and varied  the donor mass, orbital period and the
value  of  $\beta$   in  order  to  account  for   the  main  observed
characteristics  (masses of  the  components, orbital  period and,  if
available, the evolutionary status of the WD) of each system.

Let us now revisit this problem employing the set of models we present
in this  paper. Here, for  the cases of  PSR~J1713+0747, PSR~B1855+09,
and PSR~J1909-3744  we do  not try  to perform a  detailed fit  to the
observed data  as done in  the aforementioned papers but  only bracket
plausible  solutions.   Some  of  them  are possible  because  of  the
relaxation   of   the   initial    canonical   NS   mass   value.   In
Table~\ref{table:psr_fitting} we list, for  given initial donor and NS
masses,  the initial  period interval  for which  we  expect plausible
solutions for the observed parameters of the quoted systems. We remind
the reader  that these results are  restricted to the  case of $\beta=
0.5$ and $\alpha=1$.

\begin{centering}
\begin{table}
\caption{\label{table:psr_fitting}  Some tentative  initial conditions
for  the  systems  presented in  Table~\ref{table:shapiro-delay-data},
deduced  from  the  results  given in  Table~\ref{table:results}.  The
correct solution  for each  system should fall  near the  initial mass
values  and  inside  the  period  intervals  listed  below.   Here  we
considered  donor stars  with solar  metallicity and  for  the orbital
evolution we set  $\beta= 0.5$ and $\alpha= 1$. From  left to right we
list  the system  (pulsar)  name, the  plausible  interval of  initial
orbital  periods, and  the initial  masses  for the  normal donor  and
accreting NS.}
\begin{tabular}{|c|c|c|c|}
\hline 
\hline
Name & $P_i$ & $M_i$ & $(M_{NS})_i$  \\
     & $[d]$ & $[M_{\odot}]$ & $[M_{\odot}]$ \\
\hline
PSR~J1713+0747 & 6.00 - 12.00 & 1.00 & 1.00 \\
\hline
PSR~B1855+09   & 1.00 - 1.50  & 3.00 & 1.20 \\
\hline
PSR~J1909-3744 & 1.00 - 1.50  & 2.50 & 1.00 \\
\hline \hline
\end{tabular} 
\end{table} 
\end{centering}

For  the case  of  PSR~J1713+0747 here  we  do not  find any  solution
corresponding  to the  case  of $(M_{NS})_i=  1.4\;  M_{\odot}$ as  in
Benvenuto et al. (2006). In that case we found adequate configurations
for $\beta \lesssim 0.1$ but  here, after RLOF episodes the NS becomes
too   massive.   For   the   case  of   the   best  observed   system,
PSR~J1909-3744,  we  also find  plausible  solutions  but  only for  a
$(M_{NS})_i$ value well below 1.4~\msol.

\begin{figure}   
\epsfysize=300pt   
\epsfbox{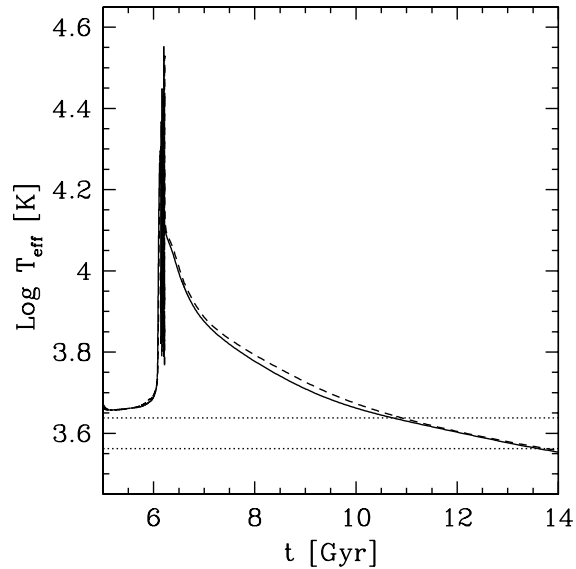} 
\caption{\label{fig:tefft} The evolution  of the effective temperature
corresponding  to  the  donor   stars  of  the  systems  described  in
Table~\ref{table:fittingJ04374715}. Solid  (dashed) line correspond to
the case  of $(M_{NS})_i$= 1.2~\msun and  $\beta$= 0.25 ($(M_{NS})_i$=
1.0~\msun  and $\beta$=  0.50). Horizontal  dotted lines  indicate the
uncertainty in the effective temperature of the WD remnant. Evidently,
both  objects  have a  very  similar  behaviour,  and have  acceptable
effective temperatures at an age interval of 10~-~13~Gyr.}
\end{figure}

Let us  perform a deeper  analysis for the PSR~J0437-4715  system. For
this case  we compute further evolutionary sequences,  not included in
Table~\ref{table:results}, for which we  allow for different values of
$\beta$ (although we still consider $\alpha=1$). We find two plausible
solutions  (see Table~\ref{table:fittingJ04374715})  that  account for
the  main  observed  characteristic   of  the  system.  Both  of  them
correspond  to an  initial donor  mass  of 1.25~\msun  and an  initial
period  of 1~day. Regarding  $(M_{NS})_i$ and  $\beta$ the  values are
1.2~\msun and  0.25 {\it or}  1.0~\msun and 0.50  respectively.  These
binary  systems provide correct  masses (both  values fall  inside the
corresponding   error   bars)   and   a   very   approximate   orbital
period\footnote{Trying to  fit the  orbital period more  accurately to
the   observed   value   does   not  change   the   presented   values
significantly. Thus,  we do not perform  a fine tuning  of the orbital
period.}. Let us compare the effective temperature of the computed WDs
with the observed value of  $T_{eff}= 4000 \pm 350$~K (Bell, Bailes \&
Bessell 1993). In Fig.~(\ref{fig:tefft})  we show the evolution of the
effective temperature of the donor  star for both systems described in
Table~\ref{table:fittingJ04374715}  together with the  observed values
interval.   We find  it  possible for  the  WD to  evolve to  observed
conditions within a time interval of 10~-~13~Gyr, shorter than (but of
the order of) the age of the Universe. Remarkably, this corresponds to
4~-~7~Gyr  after  RLOF episodes,  in  nice  agreement  with the  usual
expectation  that  this should  be  comparable  to the  characteristic
timescale  of   pulsar  rotation  braking\footnote{For   the  case  of
PSR~J0437-4715  the  period  derivative  is $\dot{P}=  5.72906  \times
10^{-20}$;  see,  e.g.,  van   Straten  et  al.  (2001).}  $\tau=  0.5
P/\dot{P}\simeq$5~Gyr observed for  PSR~J0437-4715. It is interesting to
notice     that     the      viable     solutions     presented     in
Table~\ref{table:fittingJ04374715} correspond to very different values
of $(M_{NS})_i$ and $\beta$. The main difference between these evolved
systems is the  final value of $M_{NS}$ but,  unfortunately, the large
uncertainty in  the determination of $M_{NS}$ inhibits  us to restrict
the space of parameters any further.

\begin{centering}
\begin{table*}
\caption{\label{table:fittingJ04374715}    Some    possible    initial
conditions  for the  system containing  PSR~J0437-4715 and  their main
characteristics after evolution. Both systems correspond to an initial
orbital  period   of  $P_i$=1~day  and   an  initial  donor   mass  of
1.25~\msol.  From  left to  right  we list  the  initial  mass of  the
accreting NS, the  value of $\beta$, the age and  luminosity of the WD
when its  effective temperature is $T_{eff}= 4000$~K,  the final donor
and NS  masses, and the  final orbital period. For  further discussion
see main text.}
\begin{tabular}{|c|c|c|c|c|c|c|}
\hline \hline
$(M_{NS})_i$  & $\beta$ & $t$ & $Log(L/L_{\odot})$ & $M_{WD}$ & $M_{NS}$  & $P$ \\
$[M_{\odot}]$ & & $[Gyr]$ &  & $[M_{\odot}]$ & $[M_{\odot}]$ & $[d]$ \\ 
\hline 1.20 & 0.25 & 12.081 & -3.95 & 0.2235 & 1.4563 & 5.638 \\
\hline 1.00 & 0.50 & 12.171 & -3.93 & 0.2222 & 1.5131 & 5.059 \\
\hline \hline 
\end{tabular} 
\end{table*} 
\end{centering}

The situation is more promising for the case of PSR~J1909-3744 system,
whose    mass   determinations    are   far    more    accurate   (see
Table~\ref{table:shapiro-delay-data}).   In  principle,   this  system
offers an excellent opportunity to determine the initial configuration
more accurately and even to  find {\it mean} values for the parameters
$\alpha$ and $\beta$.

In  this  Section  we   have  described  possible  solutions  for  the
primordial configuration of binary systems that evolved to account for
the  best  observed  MSP~-~WD  pairs.  Performing  a  deeper  analysis
considering variation of all parameters of the calculations (masses of
the components, orbital  period, $\beta$ and $\alpha$) is  a very time
consuming  exercise.  We  shall  defer  such  analysis  for  a  future
publication. In any case, we  consider that the results presented here
justify an effort in such direction.

\section{Discussion and Conclusions} \label{sec:discu}

In  this paper  we  perform  a set  of  binary evolution  calculations
assuming  an initial  configuration  of a  normal, solar  composition,
donor star in orbit together with  a neutron star (NS). In doing so we
consider a variety of values for  the masses of the donor and NS stars
as well as  for the initial orbital period.  These values are selected
in order to  consider systems that evolve to  ultra-compact systems or
to millisecond pulsar~-~helium white dwarf pairs (MSP~-~HeWD). In most
of the calculations we considered  that the NS accretes, at most, half
of the  matter lost by  the donor star  and that the  material ejected
from  the pair  carries  away  the specific  angular  momentum of  the
NS.  While  one of  the  main reasons  for  constructing  this set  of
calculation is  to provide  a reference frame  to analyse  the initial
configurations of  the best  observed WD~-~MSP systems,  in particular
those for which it has been possible to detect the Shapiro delay, here
we pay special attention in testing the dependence of the evolution of
these binary systems with the initial NS mass value. Also we study the
relation between  the final  orbital period and  the mass of  the HeWD
remnant.

We find that the evolution of  systems with a given orbital period and
initial mass of the normal donor  star heavily depends on the value of
the NS  mass.  For  example, we  find cases for  which, while  with an
initially  light   NS  the   system  evolves  to   an  ultra~-~compact
configuration,  if the  NS is  more massive  it gives  rise to  a well
detached  HeWD~-~NS pair. Also,  as expected,  we find  divergent mass
transfer rates (a common envelope  episode) especially for the case of
initially light NSs.

Our  calculations  show  that  the final  orbital  period~-~HeWD  mass
relation  is insensitive  to the  initial  NS mass  value, as  already
claimed  by  Rappaport et  al.   (1995).  In  any  case  we find  some
systematic departure  from the  relation proposed by  them, especially
for the case  of low mass HeWDs ($M_{WD}  < 0.25$~\msol).  This occurs
because for the  systems that give rise to such  objects, the onset of
the initial mass transfer episode occurs before the star becomes a red
giant (as  assumed in Rappaport  et al.  1995).   The best fit  to our
results   corresponds   to   Eq.~(\ref{eq:nuestrofit}).    Among   the
period~-~WD~mass relations available in the literature, we find a much
better  agreement of  our results  with  that presented  by Tauris  \&
Savonije (1999).

Employing the  set of evolutionary  sequences given in this  paper, we
also  present  preliminary  indications  of the  interval  of  initial
periods, for fixed donor and NS initial masses, inside which there are
plausible  initial  configuration for  the  binary  systems listed  in
Table~\ref{table:shapiro-delay-data}.  In  particular  we explore  the
case of the PSR~J0437-4715 system, showing that there is more than one
acceptable solution.

 \bsp

\label{lastpage} 
\end{document}